# Antiferromagnetic topological insulating state in $Tb_{0.02}Bi_{1.08}Sb_{0.9}Te_2S$ single crystals

Lei Guo[1], Weiyao Zhao[2,3*], Qile Li[3,4], Meng Xu[1,5], Lei Chen[6], Abdulhakim Bake[7], Thi-Hai-Yen Vu[3,4], Yahua He[7], Yong Fang[8], David Cortie[7,9], Sung-Kwan Mo[10], Mark Edmonds[3,4], Xiaolin Wang[7], Shuai Dong[1], Julie Karel[2,3], and Ren-Kui Zheng[4†]

[1]*Department of Physics, Southeast University, Nanjing 210096, China*

[2]*Department of Materials Science & Engineering, Monash University, Clayton VIC 3800, Australia*

[3]*ARC Centre of Excellence in Future Low-Energy Electronics Technologies, Monash University, Clayton VIC 3800, Australia*

[4]*School of Physics & Astronomy, Monash University, Clayton VIC 3800, Australia*

[5]*College of Science, Hohai University, Nanjing 211189, China*

[6]*School of Physics and Materials Science, Guangzhou University, Guangzhou 510006, China*

[7] *Institute for Superconducting and Electronic Materials, & ARC Centre of Excellence in Future Low-Energy Electronics Technologies, Innovation Campus, University of Wollongong NSW 2500, Australia*

[8] *Jiangsu Laboratory of Advanced Functional Materials, School of Electronic and Information Engineering, Changshu Institute of Technology, Changshu 215500, China*

[9] *Australia's Nuclear Science and Technology Organisation, New Illawarra Rd, Lucas Heights NSW 2234, Australia*

[10] *Advanced Light Source, Lawrence Berkeley National Laboratory, Berkeley, CA, 94720 USA*

**ABSTRACT** Topological insulators are emerging materials with insulating bulk and symmetry protected nontrivial surface states. One of the most fascinating transport behaviors in a topological insulator is the quantized anomalous Hall insulator, which has been observed in magnetic-topological-insulator-based devices. In this work, we report a successful doping of rare earth element Tb into $Bi_{1.08}Sb_{0.9}Te_2S$ topological insulator single crystals, in which the Tb moments are antiferromagnetically ordered below ~ 10 K. Benefiting from the in-bulk-gap Fermi level, transport behavior dominant by the topological surface states is observed below ~

---

* weiyao.zhao@monash.edu
† zrk@ustc.edu

150 K. At low temperatures, strong Shubnikov-de Haas oscillations are observed, which exhibit 2D-like behavior. The topological insulator with long range magnetic ordering in rare earth doped $Bi_{1.08}Sb_{0.9}Te_2S$ single crystal provides an ideal platform for quantum transport studies and potential applications.

## 1. Introduction

The concept of topological classification in condensed matter physics has led to the discovery of various exotic topological materials with nontrivial electronic band structure. Among these topological materials, the topological insulator (TI), which has gapped bulk states and gapless boundary states, has attracted research interest on its band structure, electronic transport, magnetic and optical properties. The most known TIs in 3D form are the $Bi_2Se_3$-family of compounds ($Sb_2Te_3$, $Bi_2Se_3$, $Bi_2Te_3$ and their alloys), predicted[1] and verified[2] in 2009, which have a single Dirac-cone shape topological surface state (TSS) in their bulk gap. The TSS in these compounds is protected by the bulk symmetry, and therefore is robust against non-magnetic bulk defects. Therefore, the 3DTI crystals are an ideal platform to study the Dirac fermion related physics, e.g., the nontrivial Berry phase[2-5], leading to half integer quantum Hall effect[6]; Majorana zero modes[7,8] and quantum anomalous Hall (QAH) insulating states[9,10]. These novel phenomena are the foundation to support non-Abelian statistics for quantum computing; as well as dissipationless edge modes for low-energy and high-efficiency electronics.

Since the quantum transport behavior is related to the linear-dispersed TSS, it is very important to tune the Fermi level into bulk gap, to reduce the nonrelativistic bulk carrier density. The $Bi_2Se_3$-family of compounds share the same tetradymite crystal structure, which belong to the $R\bar{3}m$ space group, with quintuple (Se/Te-Sb/Bi-Se/Te-Sb/Bi-Se/Te) layers stacked along the $c$ axis via the van der Waals interaction. As a result, the defect engineering between $p$-type $Sb_2Te_3$ and $n$-type $Bi_2Se_3$, $Bi_2Te_3$ strategy was popular, which leads to good bulk-insulating TI candidates like $Bi_{2-x}Sb_xTe_{3-y}Se_y$[11,12]. In 2016, an alternative bulk-insulating TI compound Sn doped $Bi_{1.1}Sb_{0.9}Te_2S$ with excellent properties was reported[13], in which the sulfur atoms occupy the middle of quintuple layers, and successfully reduce the Bi-Te antisite defect density. The bulk-insulating TIs benefit electronic transport studies significantly, in which the quantum Hall effect of 2D Dirac TSSs was observed without any leakage through bulk channels[6,14,15].

Moreover, it is reported that magnetic dopants will open a gap at the Dirac point on the gapless TSSs, which is the key of realizing QAH effect in a 3D TI[16]. Motivated by this, ferromagnetism was introduced into the TI system, either by magnetically doped TI compound, or by constructing TI/ferromagnet heterostructures[17,18]. In the magnetic doped TIs, the angle-resolved photoemission spectroscopy (ARPES) experiments show the gap opening at Dirac point, which further verifies the QAH insulating states[19,20]. Recently, the QAH effect was also reported in intrinsic antiferromagnetic TI $MnBi_2Te_4$ based devices, in which the odd-layer thin flakes show ferromagnetism due to the interlayer antiferromagnetic coupling[21].

Although the crystal growth studies of either a bulk-insulating TI, or a magnetically doped TI are very popular in the literature, studies of the successful magnetic doping of a bulk-insulating TI are rare. A good example is the Fe-doped $BiSbTe_2Se$, in which the scanning tunneling microscopy and spectroscopy measurements revealed the TSS with a ~ 12 meV gap opened at the Dirac point[22]. On the other hand, rare earth elements possess larger magnetic moments, which could open a larger band gap on the TSS, and potentially increase the survival temperature of QAH insulating states, e.g., in Dy doped $Bi_2Te_3$ thin films the Dirac gap is ~ 85 meV at room temperature[23]. Here we report the growth of a rare earth Tb doped bulk-insulating TI $Bi_{1.1}Sb_{0.9}Te_2S$ (TbBSTS) single crystal by modified Bridgeman method. The Tb dopants are uniformly distributed in the single crystal and antiferromagnetically ordered at ~ 10 K. With Tb dopants, the single crystal still shows bulk-insulating TI behavior, which has an insulator-to-metal transition upon cooling, illustrating the TSSs' dominance at low temperatures. The Shubnikov-de Haas oscillations with nontrivial Berry phase are observed at low temperatures, which shows 2D behavior in angular dependent measurements. All the evidence demonstrates that Tb:BSTS is an excellent bulk-insulating TI with antiferromagnetic ordering, which will be an ideal platform for TI-based device designs.

## 2. Experiments

The single crystals of Tb:BSTS were grown using a slow-cooling method[4,24]. Briefly, 1) the high impurity powders of Tb, Bi, Sb, Te and S in stoichiometric ratio ($Tb_{0.02}Bi_{1.08}Sb_{0.9}Te_2S$) are sealed in silica tube under vacuum, 2) the mixture was heated to 1100 °C, then cooled to 500 °C with 2 °C/h speed, 3) large pieces of single crystal (5 × 5× 1 $mm^3$) were obtained in the ingot, in which one piece with shiny surface is employed in the magnetic and transport study. The orientation of the single crystal is verified by X-ray diffraction (XRD) patterns which were measured using the Rigaku SmartLab X-ray diffractometer. The chemical compositions and element mapping were measured using an energy dispersive X-ray spectrometer (EDS, Oxford Aztec X-Max80) installed on a Zeiss Supra 55 scanning electron microscope. Electronic transport and magnetic properties were measured using a physical property measurement system (PPMS, DynaCool-14, Quantum Design).

## 3. Results and Discussion

After single crystal growth, shiny single crystal flakes with large in-plane size can be exfoliated from the ingot. With 1/3 S doped at Te site, the S prefers the center of the quintuple layer[25] (as shown in Fig. 1a), which significantly reduces the Te-Bi antisite defects[13]. The sharp comb-like XRD patterns of (00$l$) peaks are obtained from the XRD experiments. The XRD $\omega$-scan around (0018) peak yields a rocking curve with full-width at half-maximum (FWHM) value of ~0.01 °. The EDS mapping of different elements was conducted on a fresh-cleaved single crystal surface. Over a sufficiently large area, both major elements and Tb dopants show a uniform distribution without segregation. Moreover, the EDS shows the Tb doping level ~ 0.3 %, which is slightly lower than the nominal doping level. Due to the low doping concentration, it is very important to further confirm the phase purity before further transport study. As shown in Supplementary Materials[26] Figs. S1&2, the X-ray diffraction of fine powder grounded from multiple high quality crystal pieces, zoom-in energy-dispersive spectra suggests that the Tb are probably doped into crystal lattice. The $Bi_{1.1}Sb_{0.9}Te_2S$ related compound show excellent bulk-insulating TI property, e.g., the transport band gap obtained from thermal activation fitting is 0.1 – 0.3 eV[4,13], and sharp surface bands with clear Dirac point are observed in ARPES[13]. Therefore, the transition from insulating-to-metallic behavior can be observed in the temperature dependent resistivity curve upon cooling, due to the gapless TSS being more pronounced at low temperatures. As shown in Fig. 1e, the resistivity increases exponentially with cooling, and it reaches a maximum value (13.1 Ω·cm) at ~ 100 K. The exponentially increasing part at high temperatures can be fitted with Arrhenius equation $ln\rho(T) = ln\rho_0 +$

$E_g/2k_BT$, in which the $\rho_0$ is residual resistance at 0 K, $E_g$ is the band gap, $k_B$ is the Boltzmann's constant, and $T$ is temperature, shown in Fig. 1f. The obtained bulk band gap in transport measurements is ~ 236 meV. Upon further cooling, the resistivity decreases (to ~ 10.7 Ω·cm at 2 K), which implies the dominance of gapless TSS.

To explore the magnetic ordering of the Tb dopants in the TI crystal, temperature dependent magnetization was measured via a vibration sample magnetometer (VSM) equipped on PPMS, as shown in Fig. 1g. During the measurements, the magnetization data were taken in the heating process with a small magnetic field (0.02 T) applied along the *c* axis in both zero-field-cooling (ZFC, no fields applied during cooling) and field-cooling (FC, 0.02 T field was also applied during cooling from room temperature) mode. With cooling, the magnetic susceptibility increases exponentially first, and reaches a maximum value at ~ 10 K. then decreases. The temperature dependent magnetic susceptibility $\chi(T)$ shows a clear antiferromagnetic feature with a Néel temperature $T_N$ ~ 10 K, which is a result of the Tb doping. The antiferromagnetic ordering was also reported in Ho, Gd doped TI crystals or thin films [27-29]. Note that, the ZFC curve and FC curve are coincide with each other, which demonstrates no ferromagnetic coupling between Tb dopants, nor ferromagnetic impurities in the obtained single crystals. The high temperature $\chi(T)$ curve follows Curie-Weiss law $\chi = C/(T + \theta_p)$, in which the $C$ is the Curie constant, and $\theta_p$ is the Weiss constant. The fitting of the $\chi(T)$ data is shown in Fig. 1h, and the magnetic moment (convert to each Tb using the nominated ratio) calculated from Curie constant is $\mu_{Tb}$ ~ 7.8 $\mu_B$, employing $C = (N_A \mu_{Tb}^2)/3k_B$, and $N_A$ is the Avogadro number. Further, the magnetization *vs.* magnetic field (M(H)) measurement is also conducted in the 2 – 50 K range, with applied magnetic fields up to 14 T. At 2 K, the magnetization increases with applied field linearly below 2 T, which is the typical response of an antiferromagnetic material. However, in the 2-3 T region, the system shows a spin-flop-like transition, after which the magnetization increases with magnetic field faster than the low-field region. Above 10 T, the magnetization shows a sign of saturation, with the magnetization values reaches ~ 8 $\mu_B$/Tb. The spin-flop transition occurs when magnetic field applied parallel to the antiferromagnetic axis (which is the hard axis), and above the transition field, the antiferromagnetic axis rotates to perpendicular to the field (easy axis). Therefore, we deduce that the Tb moments are aligned along the *c* axis of the TI crystals, which is sketched in Fig. 1a. The similar M(H) curve can be observed at 5 K, which agrees with the plateau-like $\chi(T)$ behavior below ~ 5 K in Fig. 1g. At 7.5 K, the spin-flop-like transition was suppressed, which suggests the antiferromagnetic coupling is relatively soft at this temperature, which is in the $\chi$ fast-decreasing region below $T_N$. Near $T_N$, the spin-flop transition disappears, and the M(H) curve shows easy-axis behavior. Above $T_N$, the sample shows linear a M(H) curve, as expected in the paramagnetic phase. It is worth mentioning that the similar M(H) behavior antiferromagnetic phase can be found terbium antimonide crystals. In Supplementary Material[26] (see, also, references[30,31] 30 and 31 therein) Fig S4, a *dM/dB* plot of M(H) curve at 2 K demonstrates the difference between Tb_BSTS and Tb antimonide. Even with several positive evidences, the accurate Tb position in BSTS lattice remains a mystery, which is essentially important to the magnetic and electronic structure of the Tb_BSTS crystals.

The magnetotransport measurements of TbBSTS crystal are important to understand the electronic band structure of this antiferromagnetic TI material, which are shown in Fig. 2. In Fig. 2a, the magnetoresistance ($MR = \frac{\rho(B)-\rho(0)}{\rho(0)} \times 100\%$) at different temperatures are stacked with constants as offset, to better demonstrate the details of each curve. At 2 K, the MR increases with applied fields first, then shows significant oscillation behavior above 2 T. The oscillations are denoted as Shubnikov-de Haas (SdH) oscillations, which describes the Landau quantization effect of the Fermi surface[32]. To obtain oscillation patterns, the negative second derivative curves were calculated to remove the background and plotted in Fig. 2b ($-cos''(x) = cos(x)$). The SdH oscillation amplitude damps with heating, and is still observable at ~ 50 K, which is similar to the report in similar compounds[4]. With further heating, the quantum oscillation vanishes. Below 75 K, the background MR (without oscillation patterns) increases with magnetic field monotonically. In the low field region, the MR increases rapidly with applied field, showing a cusp shape curve near 0 T, which is due to the weak antilocalization behavior in TI materials[33,34]. Above ~ 1 T, the MR increases linearly with magnetic field, which is possibly due to the mobility fluctuation[35,36] induced by the defects. Above ~ 7 T, the MR show further linear increase beyond the last oscillation peak, consistent with the quantum linear MR model[37]. Based on the RT curve in Fig. 1e, the TSS dominantes the transport behavior below 100 K, where the SdH oscillations and linear-like MR are observed. At 100 K and above, the bulk conducting channels are more pronounced, which provide the negative MR due to paramagnetic Tb impurities.

Since the TSS's contribution to transport property decreases with heating, one may expect the competition of dominance at a certain temperature in both MR and the Hall effect. The Hall effect curves at different temperatures are shown in Fig. 2d, in which the curves nearly coincide with each other at and below 50 K, and evolve with heating significantly above 100 K. It is worth revisiting the band structure of the $Bi_{1.1}Sb_{0.9}Te_2S$ crystal, which possesses a bulk band gap of ~ 0.34 eV, and a Dirac point of the TSS located 0.12 eV above the valence bands[13]. Due to the RT fitting in Fig. 1f, the thermal activation energy $E_a \sim E_g/2 \sim 0.12$ eV, which means the $E_f$ is in the bulk band gap, and located ~0.12 eV away from bulk bands. In Fig. 2d, the low temperature Hall curve shows electron-dominant shape, and "$S$" shape, indicating the multi-band contribution, at 150 K. The multi-band fitting model can be written as:

$$\rho_{xy} = \frac{B}{e} \frac{(n_h \mu_h^2 - n_e \mu_e^2) + (n_h - n_e)\mu_h^2 \mu_e^2 B^2}{(n_h \mu_h - n_e \mu_e)^2 + (n_h - n_e)^2 \mu_h^2 \mu_e^2 B^2} \qquad 1$$

where $e$ is the elemental charge, $n_h, n_e, \mu_h, \mu_e$ are density and mobility of hole carrier's and electron carriers, respectively. The fitting curve is shown in Fig. 2e, from which the $n_h, n_e$ are $1.28 \times 10^{18}$ and $0.69 \times 10^{18}$ cm$^{-3}$; $\mu_h, \mu_e$ are $1.1 \times 10^5$ and 29 cm$^2$/Vs. The same fitting model is employed to calculate the carrier density and Hall mobility at all temperatures, which are shown in Fig. 1f&g. Since the TbBSTS show electron dominant behavior at low temperatures, and hole dominant behavior near room temperature, the Fermi level is thus deduced to be ~ 0.12 eV above the bulk valence bands.

The SdH patterns shown in Fig. 2b are plotted in two separated parts, to emphasize the high-field patterns (left part on $1/B$ axis). The multi-frequency SdH oscillations implies multiple Fermi pockets in the measured specimen, which indicates the Dirac cone is offset in the top and bottom surface TSSs in a bulk-insulating 3DTI[14,38]. The Landau quantization modulates the conductivity ($\sigma_{xx}$) in magnetic fields, which shows a peak when the Fermi level is located in the Landau levels. Further, one may learn the fermiology from the SdH oscillations using the using Lifshitz-Kosevich (LK) formula, with the Berry phase being considered:

$$\frac{\Delta\rho}{\rho(0)} = \frac{5}{2}\left(\frac{B}{2F}\right)^{\frac{1}{2}} R_T R_D R_S \cos\left(2\pi\left(\frac{F}{B} + \gamma - \delta\right)\right) \qquad 2$$

where $R_T = \alpha T m^*/B\sinh(\alpha T m^*/B)$, $R_D = \exp(-\alpha T_D m^*/B)$, and $R_S = \cos(\alpha g m^*/2)$. Here, $m^*$ is the ratio of effective cyclotron mass to the free electron mass $m_e$; $g$ is the $g$-factor; $T_D$ is the Dingle temperature; and $\alpha = (2\pi^2 k_B m_e)/\hbar e$, where $k_B$ is Boltzmann constant, $\hbar$ is the reduced Planck constant, and $e$ is the elementary charge. The frequency of oscillation patterns in Fig. 2b can be obtained from their fast Fourier transform (FFT) spectra, as shown in Fig. 2c: $F_\alpha$ = 5 T, $F_\beta$ = 44.4 T. According to the Onsager-Lifshitz equation, the frequency of quantum oscillations, $F = (\varphi_0/2\pi^2)A_F$, where $A_F$ is the extremal cross-sectional area of the Fermi surface perpendicular to the magnetic field, and $\varphi_0$ is the magnetic flux quantum. Therefore, the cross-section area $A_F$ related to the two Fermi pockets are 4.8×10$^{-4}$ and 4.2×10$^{-3}$ Å$^{-2}$, respectively. Using $A_F = \pi k_F^2$, the Fermi wave vector $k_F$ are 0.012 and 0.037 Å$^{-1}$ for $F_\alpha$ and $F_\beta$, respectively. In Fig. 2b&c, one can see that the SdH oscillation patterns are shifting with temperature change, e.g., the patterns are identical in the 2 – 10 K region, however shift to low frequency at 25 K and higher temperatures. The frequency red shifting means the Fermi pockets are shrinking with heating above 10 K, which is possibly related to the magnetic ordering of the Tb dopants. Another interesting information one can obtained from the SdH oscillations is Berry phase, which is the phase shift in the cosine term of the LK formula. The phase factor is $\gamma - \delta$, in which $\delta = 0$ for 2D Fermi pockets, and ±1/8 for 3D Fermi pockets, $\gamma = 1/2 - \Phi_B/2\pi$, where $\Phi_B$ is the Berry phase. In Tb_BSTS system, the Berry phase of SdH oscillations can be obtained by extrapolating the Landau level (LL) index $n$ to the extreme field limit (1/B→0) in the Landau fan diagram, as shown in the Inset of Fig. 2(c). Our resistivity measurements show that $\rho_{xx} \ll \rho_{xy}$ in the crystals, which means that $\sigma_{xx}$ is in phase with $\rho_{xx}$, we assign the maximum of the oscillations as half integer Landau index, the minimum of the oscillations as integer Landau index, respectively, and linearly fitted the data. In this case, the 1/2 phase shift in $\gamma$ has been considered to in the Landau indices assignment, which means that with $\pi$ Berry phase, the intercept is 0.5. The LL index fittings show 0.32±0.21 for $F_\alpha$ and 0.41±0.18 for $F_\beta$, which relates to 0.64±0.42 $\pi$ and 0.82±0.36 Berry phase for each pocket. The Berry phase in a massive Dirac fermion system can be tuned continuously from $\pi$ to 0 with Fermi surface shifting from far away to into the massive gap[20]. The off-π Berry phase values may indicate the massive Dirac fermion model, however, due to the large error bar, further evidences are desired to confirm this point. In Supplementary Materials[26] Fig. S3, a quick ARPES image taken at ~7 K near the Fermi surface also shows the gapped Dirac bands, which further supports this point.

Further, we conducted rotation angle measurements in TbBSTS crystal, as demonstrated in Fig. 3. Note that, during rotation from the $c$ axis (0 °) to the ab plane (90 °), the magnetic field is always perpendicular to the current direction, to eliminate the influence of Lorenz force. In Fig. 3a, the linear-like MR effect is robust against rotation, however, the SdH oscillation patterns on the MR curve shifts and weakens obviously. Therefore, the oscillation patterns were subtracted by calculated the negative second derivative, as shown in Fig. 3b, in which the pattern shifting is better demonstrated. The FFT spectra of these oscillation patterns are also calculated and shown in Fig. 3c, in which the $F_\beta$ region is plotted in zoom-in mode. The frequencies in the FFT spectra are summarized in Fig. 3d. For a 2D Fermi surface, the angular-dependent SdH frequency $F(\theta)$ increases in an inverse cosine rule: $F(\theta) = F(0)/\cos(\theta)$, which is shown as fitting lines in Fig. 3d. As expected, the oscillations in Tb:BSTS follows the 2D rotation rule during rotation, which confirms that the 2D TSSs contribute to the SdH oscillations.

## 4. Conclusion

In the view of single crystal growth, rare earth elements are better dopants than transition metals, for the similar radii and the same valence state. Therefore, high quality single crystal with large size, high carrier mobility and various magnetic ordering can be expected in rare earth doped TI samples[24,28,39,40]. Here, we show the successful doping of Tb in a bulk-insulating 3D TI $Bi_{1.1}Sb_{0.9}Te_2S$, which shows antiferromagnetic ordering between layers below $T_N$ ~ 10 K. The topological insulating behavior with the Fermi level located in bulk gap is preserved with Tb doping, e.g., the bulk band gap in transport measurements is ~ 0.236 eV, an insulator-to-metal transition occurs near 100 K, below which the metallic topological surface states dominate the transport behavior. The strong SdH oscillations are observed at low temperatures, which arise from the 2D topological surface states. Due to the local impurity level difference, the Dirac cone offset between the top and bottom surface exists, and contributes to multi-frequency SdH oscillations. The low-frequency oscillation shows a quantum limit of 5 T, which suggest the quantized Hall conductance could be observed in thin-flake-based devices. The intrinsic antiferromagnetic TI MnBi$_2$Te$_4$ has attracted great attention for the QAH effect[21] and axion insulating states[41,42], which suggests the importance of antiferromagnetism in TI study. The bulk-insulating antiferromagnetic TI phase in TbBSTS will encourage more research on exploring the local chemical environment of doping elements, the magnetic structure, and their relationship to electronic band structure in rare earth doped TI system.

**ACKNOWLEDGEMENT** This work is supported by ARC Centre of Excellence in Future Low-Energy Electronics Technologies No. CE170100039, Australian Research Council Discovery Project DP200102477, National Natural Science Foundation of China (Grant No. 11974155, 12174039). This research used resources of the Advanced Light Source, which is a DOE Office and Science User Facility under contract no. DE-AC02-05CH11231. WZ acknowledge the supporting from AINSE.

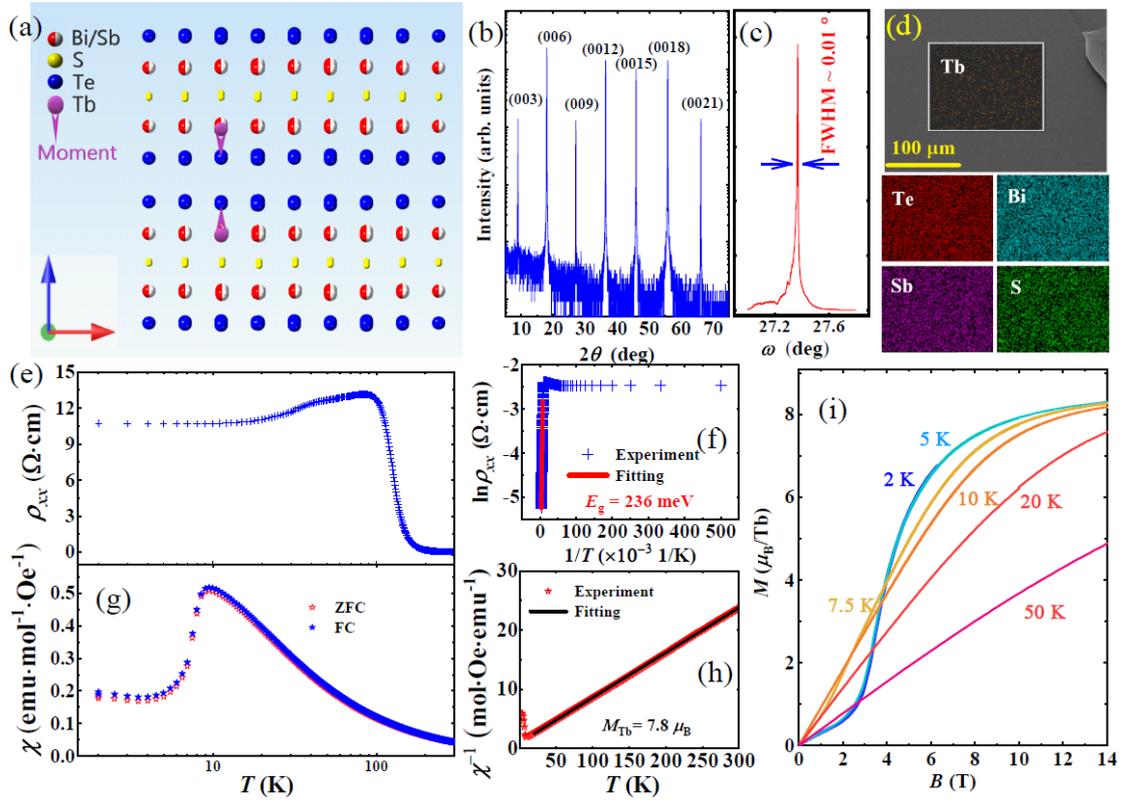

Fig. 1 The physical properties of TbBSTS single crystal sample. (a) The schematics of the crystal structure, in which the terbium's substitution of Bi/Sb site is assumed. (b) The XRD patterns of (00$l$). The X-ray rocking curve of (0018) peak is shown in Panel (c). (d) The EDS element maps were taken on a fresh-cleaved smooth area, which demonstrate the uniform distribution of all elements. The temperature dependence of resistivity (e) and magnetic susceptibility (g) are plotted as a function of temperature. To emphasize the low-temperature feature, the logarithm scale is employed here. (f) The thermal activation fitting of the insulating part of Panel (e). (h) The Curie-Weiss fitting of the paramagnetic part of Panel (g). (i) The MH curves at different temperatures from 2 K to 50 K.

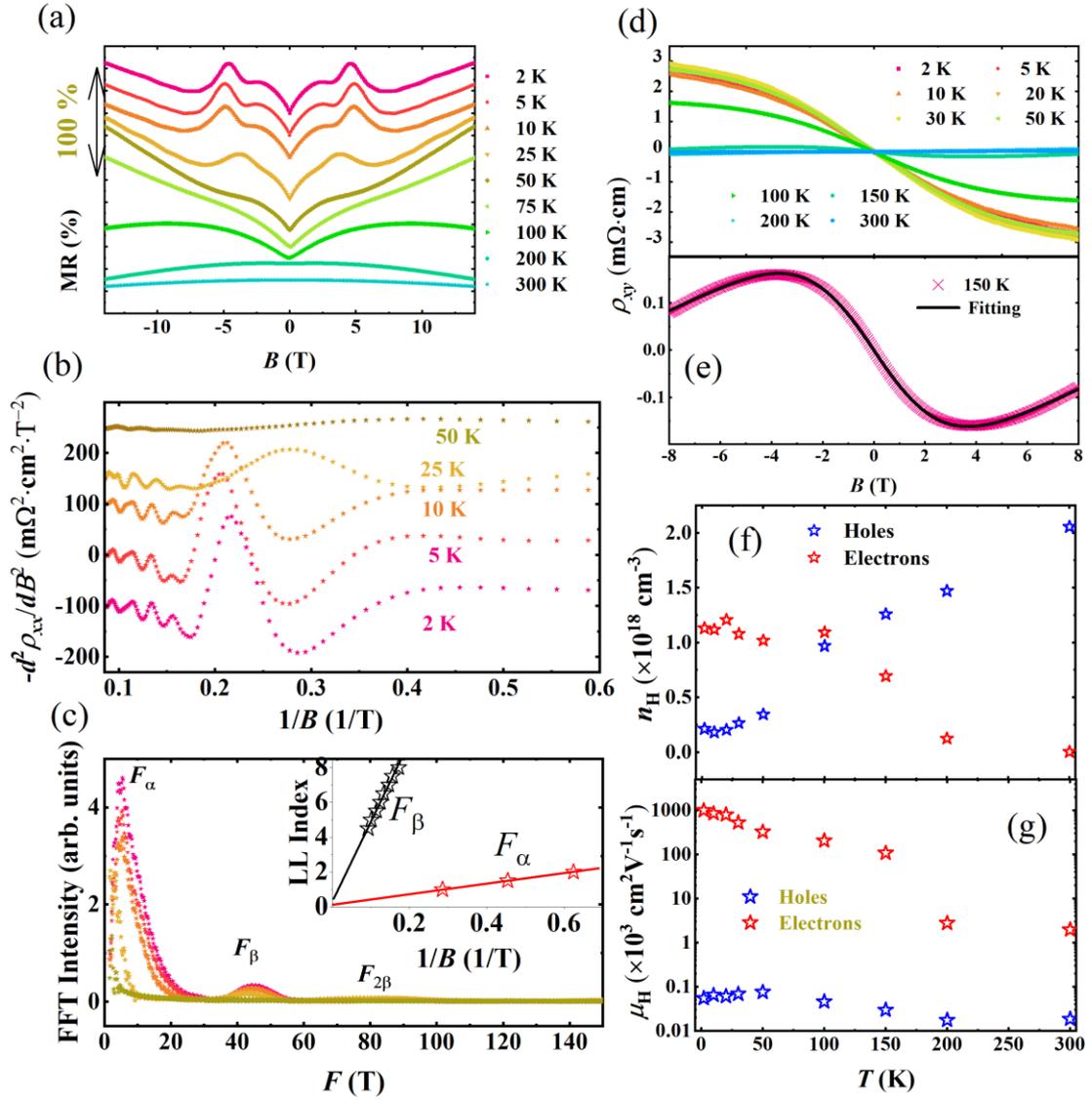

Fig. 2 The magnetotransport results of TbBSTS crystal sample. (a) The temperature dependent magnetoresistance curves are plotted in stacking mode with constants bias between different curves. (b) The SdH oscillation patterns obtained from Panel (a) are plotted with 1/$B$, the FFT patterns are shown in the Panel (c). The Inset of Panel c shows the Landau index fitting of oscillation patterns at 2 K. (d) The Hall effect are also measured at different temperatures. At ~ 150 K, the Hall curve shows clear two-carrier sign, which is fitted using multicarrier model and shown in Panel (e). The carrier's concentration and mobility of electron and holes are plotted in Panel (f) and (g), respectively.

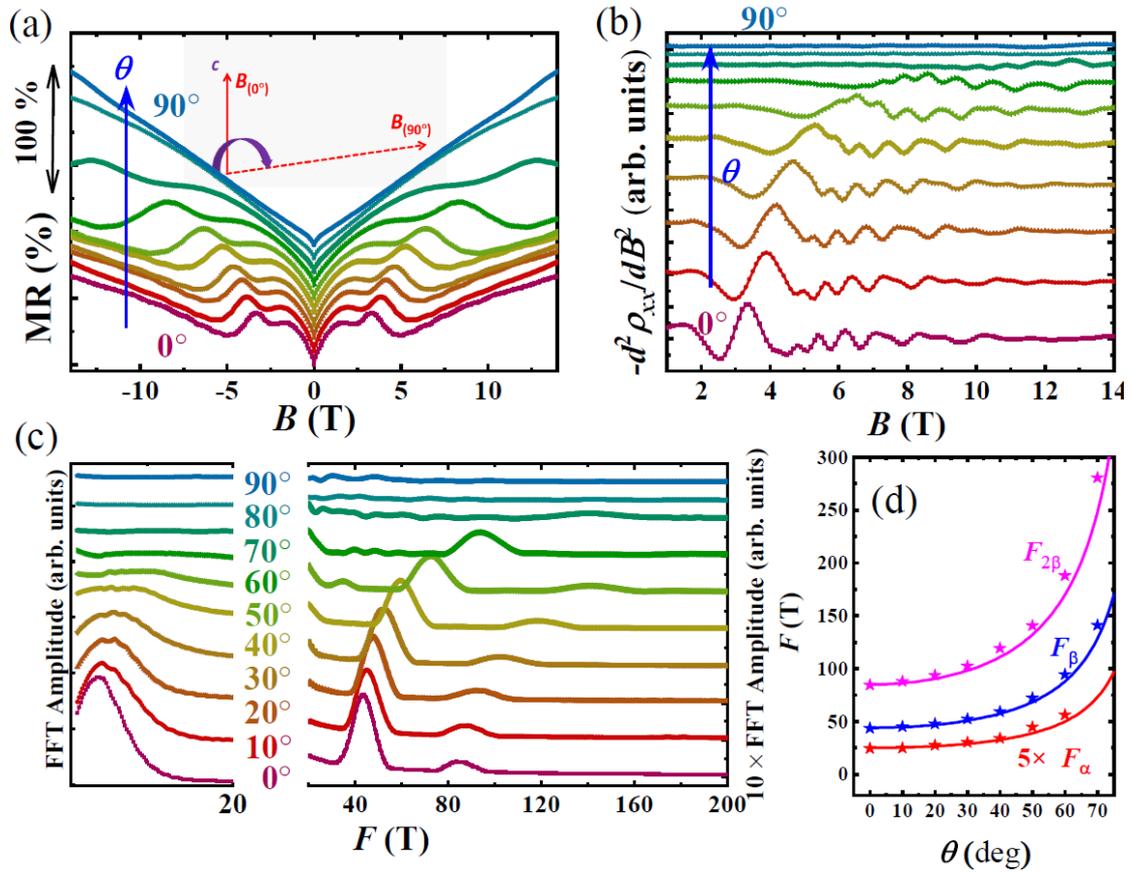

Fig. 3 The rotation angle measurements of MR curves at 2 K. (a) The MR curves are plotted in stacking mode with constant offset values among different rotation angles, in which the inset sketch shows the geometry of rotation angle. (b) The background-subtracted oscillation patterns. (c) The FFT spectra of oscillation patterns in Panel (b). (d) A summarize of the obtained FFT frequencies from Panel (c). Note that, the $F_\alpha$ is plotted in 5 times larger value, to demonstrate the details.